\documentclass[iop]{emulateapj}
\usepackage{natbib}
\usepackage{amsmath}

\newcommand{\msun}{$M_{\odot}$}
\newcommand{\mjup}{$M_{J}$}
\newcommand{\rsun}{$R_{\sun}$}
\newcommand{\ldl}{$\lambda/{\Delta}{\lambda}$}
\newcommand{\teff}{$T_\mathrm{eff}$}
\newcommand{\logg}{$\log{g}$}

\newcommand{\kms}{km~s$^{-1}$}

\newcommand{\name}{WISE~J052857.68+090104.4}
\newcommand{\namesh}{WISE~J0528+0901}
\newcommand{\teffresult}{1880$^{+150}_{-70}$}
\newcommand{\loggresult}{3.8$^{+0.2}_{-0.2}$}

\slugcomment{Accepted to ApJ 2016 Feb 8}

\shorttitle{Planetary Mass Object in 32 Ori}
\shortauthors{Burgasser et al.}

\begin{document}

\title{The First Brown Dwarf/{Planetary-Mass Object} in the 32 Orionis Group\footnote{This paper includes data gathered with the 6.5 meter Magellan Telescopes located at Las Campanas Observatory, Chile.}}

\author{Adam J. Burgasser\altaffilmark{1}, 
Mike A. Lopez\altaffilmark{1}, 
Eric E. Mamajek \altaffilmark{2}, 
Jonathan Gagn\'e\altaffilmark{3}, 
Jacqueline K. Faherty\altaffilmark{4,5,6}, 
Melisa Tallis\altaffilmark{1}, 
Caleb Choban\altaffilmark{1}, 
Ivanna Escala\altaffilmark{1}
and Christian Aganze\altaffilmark{1,7}}

\altaffiltext{1}{Department of Physics, University of California, San Diego, CA 92093, USA; aburgasser@ucsd.edu}
\altaffiltext{2}{Department of Physics and Astronomy, University of Rochester, Rochester, NY 14627-0171, USA}
\altaffiltext{3}{Institut de Recherche sur les Exoplan\`etes (iREx), Universit\'e de Montr\'eal, Département de Physique, C.P. 6128 Succ. Centre-ville, Montr\'eal, QC H3C 3J7, Canada}
\altaffiltext{4}{Department of Terrestrial Magnetism, Carnegie Institution of Washington, Washington, DC 20015, USA }
\altaffiltext{5}{Department of Astrophysics, American Museum of Natural History, Central Park West at 79th Street, New York, NY 10024, USA }
\altaffiltext{6}{Hubble Fellow }
\altaffiltext{7}{Department of Physics and Dual-Degree Engineering, Morehouse College, 830 Westview Drive S.W, Atlanta, GA 30314, USA}

\begin{abstract}
The 32 Orionis group is a co-moving group of roughly 20 young (24~Myr) M3-B5 stars 100~pc from the Sun. Here we report the discovery of its first substellar member, WISE~J052857.69+090104.2. This source was previously reported to be an M giant star based on its unusual near-infrared spectrum and lack of measureable proper motion.  We re-analyze previous data and new moderate-resolution spectroscopy from Magellan/FIRE to demonstrate that {this source} is a young near-infrared L1 brown dwarf with very low surface gravity features. Spectral model fits indicate {\teff} = {\teffresult}~K and {\logg} = {\loggresult}, consistent with a 15--22~Myr object with a mass near the deuterium-burning limit. Its sky position, estimated distance, kinematics (both proper motion and radial velocity), and spectral characteristics are all consistent with membership in 32~Orionis, {and its temperature and age imply a mass (M = 14$^{+4}_{-3}$~{\mjup}) that straddles the brown dwarf/planetary-mass object boundary.}
The source {has a somewhat red $J-W2$ color compared to other L1 dwarfs, but this is likely a low-gravity-related {temperature} offset; we find no evidence of significant} excess reddening {from a disk or cool companion} in the 3--5~$\micron$ waveband.
\end{abstract}


\section{Introduction}

Current star formation theory holds that the vast majority of stars form in clusters or groups,
although whether most come from massive star-forming regions 
or low-density associations remains a matter of ongoing debate \citep{2001ApJ...553..744A,2003ARA&A..41...57L,2010MNRAS.409L..54B,2014ApJ...791..131K}.
The origin of brown dwarfs, objects with insufficient mass to fuse hydrogen (M $\lesssim$ 0.07~{\msun}; \citealt{1962AJ.....67S.579K,1963ApJ...137.1121K,1963PThPh..30..460H}) is even more uncertain, as these sources are detectable only at relatively short distances, roughly out to the Orion Nebula Cluster 
(ONC) and Ori OB1 subgroups ($d$ $\approx$ 400~pc). Young brown dwarfs  are increasingly being found in nearby ($d$ $\lesssim$ 100~pc) sparse associations and moving groups \citep{2004ARA&A..42..685Z,2006ApJ...643.1160L,2008hsf2.book..757T,2002ApJ...575..484G,2010AJ....140..119S,2014ApJ...783..121G,2015arXiv150607712G} thanks to the infrared sensitivity and multi-epoch astrometry provided by the 2 Micron All Sky Survey (2MASS;  \citealt{2006AJ....131.1163S}) and the Wide-field Infrared Survey Explorer (WISE; \citealt{2010AJ....140.1868W}), among others.  
The organizations of these groupings vary considerably, and include
clusters, or coeval groups of gravitational bound stars; associations, or coeval groups of stars that are gravitationally unbound and will disperse over $\approx$10$^7$--10$^8$~yr; and moving groups or streams, which also have common kinematics but may or may not be coeval \citep{2004ARA&A..42..685Z,2006ApJ...643.1160L,2008hsf2.book..757T}. 
Unlike massive star clusters, nearby associations and moving groups are more widely dispersed and have fewer stars ($\lesssim10^2$ vs $10^4$ for the ONC), making the detection of members challenging. As such, the populations of the most common nearby associations remain incomplete, particularly at substellar masses.
Identification and study of brown dwarfs in these systems is essential for characterizing the birth sites and formation processes for the Galactic brown dwarf population, as well as characterizing
the influence of age and mass on atmospheric chemistry (e.g., cloud-formation; \citealt{2008ApJ...674..451B,2008ApJ...686..528L,2011ApJ...735L..39B}), 
investigating disk and planet formation and evolution as a function of primary mass (e.g., \citealt{2002ApJ...578L.141J,2005ApJ...620L..51L}), 
and identifying analogs of young exoplanetary systems (e.g., \citealt{2008Sci...322.1348M,2010ApJ...719..497L,2013A&A...553L...5D,2013AJ....145....2F,2015ApJ...804...96G}).

Recently, \citet{2013PASP..125..809T} reported the detection of a faint infrared source {\name} (hereafter {\namesh}) whose near-infrared spectrum exhibits features consistent with very low surface gravity. Due to the lack of measureable proper motion in the original WISE data, they concluded that this source is an M giant star.  In this paper, we demonstrate that this source is a young L-type brown dwarf/planetary mass object and the first substellar member of the 24~Myr-old 32 Orionis group \citep{2007IAUS..237..442M,2010AAS...21542822S,2015MNRAS.454..593B}. 
In Section~\ref{sec:obs} we present new near-infrared spectroscopic measurements of {\namesh} obtained with the Folded-port InfraRed Echellette spectrograph (FIRE; \citealt{2013PASP..125..270S}) at the Magellan Telescopes.
In Section~\ref{sec:spex_analysis} we re-analyze the Thompson et~al.\ spectrum by {comparing} to equivalent data of field and young brown dwarfs in the SpeX Prism Library (SPL, \citealt{2014arXiv1406.4887B}). This analysis demonstrates that {\namesh} is a near-clone to a previously-identified young brown dwarf in the 23~Myr-old $\beta$ Pictoris Moving Group, and we derive temperature and gravity classifications that support its low temperature and surface gravity. In Section~\ref{sec:model_fit} we compare the near-infrared spectrum to atmosphere models to determine the physical properties of {\namesh} and confirm its youth. In Section~\ref{sec:association} we examine spatial and kinematic evidence of membership in 32 Orionis. Our results are discussed in Section~\ref{sec:discussion}. 

\section{FIRE Observations}\label{sec:obs}

We obtained new moderate-resolution near-infrared spectral data of {\namesh} with Magellan/FIRE on 2013 December 12 (UT), in clear conditions and 0$\farcs$8 seeing at $J$-band. We used the cross-dispersed echellette mode and 0$\farcs$6 slit to obtain 0.8--2.45 $\mu$m spectroscopy at a resolving power {\ldl} $\approx$ 6000. Four exposures of 900~s each were obtained at different positions along the slit and at an airmass of 1.29.  This was followed by observations of the A0~V star HD~33831 ($V$ =  8.02) for telluric correction, flux calibration, {and first-order correction of wavelength-dependent slit losses}; and ThAr emission lamps for wavelength calibration. We obtained high- and low-illumination flat fields at the beginning of the night for pixel response calibration. All data were reduced using an updated version of the Interactive Data Language (IDL) pipeline FIREHOSE \citep{jonathan_gagne_2015_18775}, which is based on the MASE \citep{2009PASP..121.1409B} and SpeXTool \citep{2003PASP..115..389V,2004PASP..116..362C} packages.

\begin{figure*}[t]
\epsscale{0.8}
\plotone{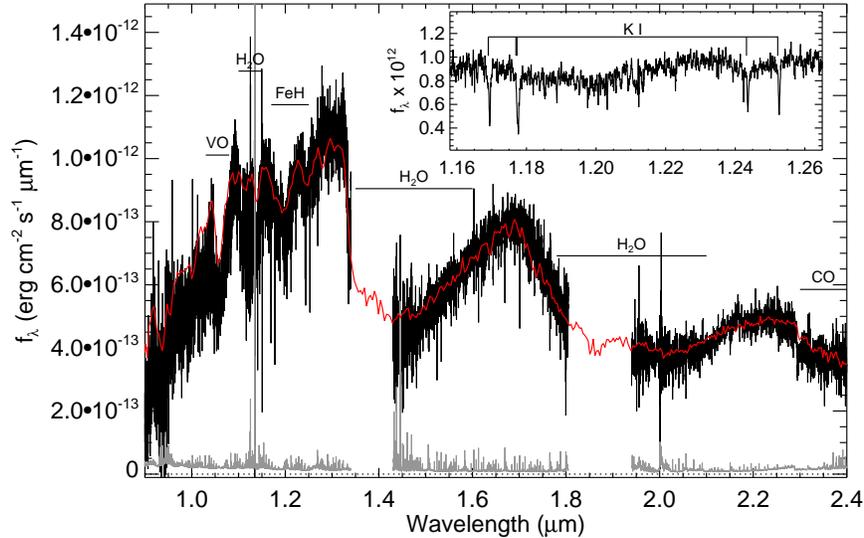}
\caption{Magellan/FIRE spectrum of {\namesh} (black line) compared to IRTF/SpeX data from \citet[red line]{2013PASP..125..809T}. Both spectra are scaled to the apparent 2MASS $K_s$ magnitude of the source. The uncertainty spectrum for the FIRE data is shown in grey. Primary molecular absorption features are labeled, while regions of strong telluric absorption have been masked.  The inset box displays the 1.160--1.265~$\micron$ region to highlight the K~I lines present in the FIRE spectrum used for RV measurement. \label{fig:fire}}
\end{figure*}

The reduced spectrum is shown in Figure~\ref{fig:fire}, compared to the SpeX spectrum of \citet{2013PASP..125..809T} which was used to guide our relative scaling of the individual FIRE orders. The FIRE spectrum appears comparatively noisier, but much of the structure seen arises from overlapping molecular transitions of H$_2$O, FeH, CO and VO present in the atmosphere of {\namesh}. The data sufficiently resolve the K~I doublets at 1.17~$\micron$ (3p$^6$4p $\rightarrow$ 3p$^6$3d) and 1.25~$\micron$ (3p$^6$4p $\rightarrow$ 3p$^6$5s), which we find to be much weaker in strength than those of field late-M and L dwarfs (equivalent widths = 3.0$\pm$0.2~{\AA} versus 5--10~{\AA} for M9--L1 dwarfs; \citealt{2003ApJ...596..561M}). We used the three lines at\footnote{Line centers are in vaccuum wavelengths and are from the NIST Atomic Line Spectral Database \citep{NIST}.} 1.1693442~${\micron}$, 1.2435700~${\micron}$ and 1.2525591~${\micron}$ to measure a heliocentric radial velocity (RV) for {\namesh} of +18$\pm$3~{\kms}, where the uncertainty includes both scatter in the line measurements and 0.8~{\kms} uncertainty in the wavelength calibration. 

\section{Re-analysis of SpeX Data}\label{sec:spex_analysis}

\begin{figure}[h]
\epsscale{0.9}
\plotone{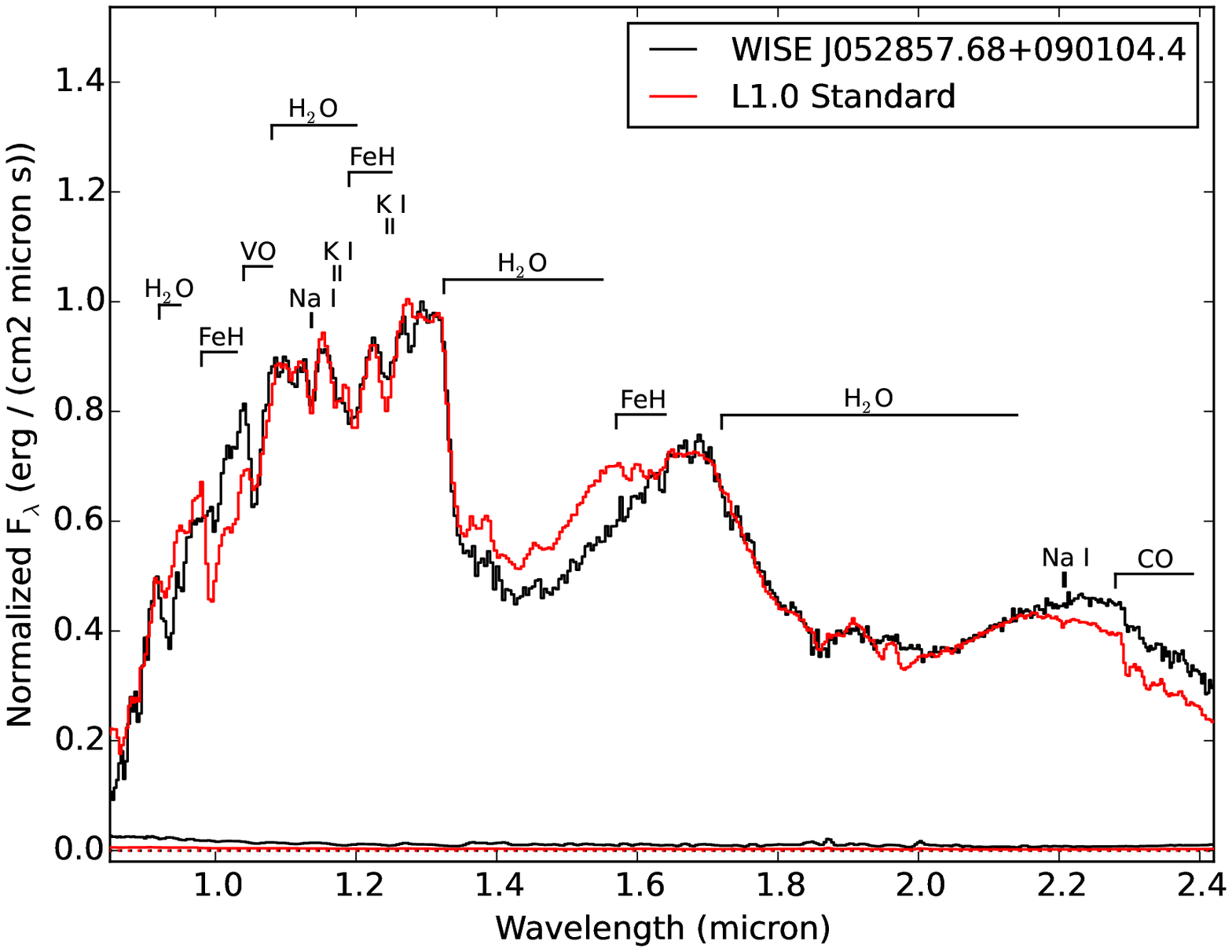}
\plotone{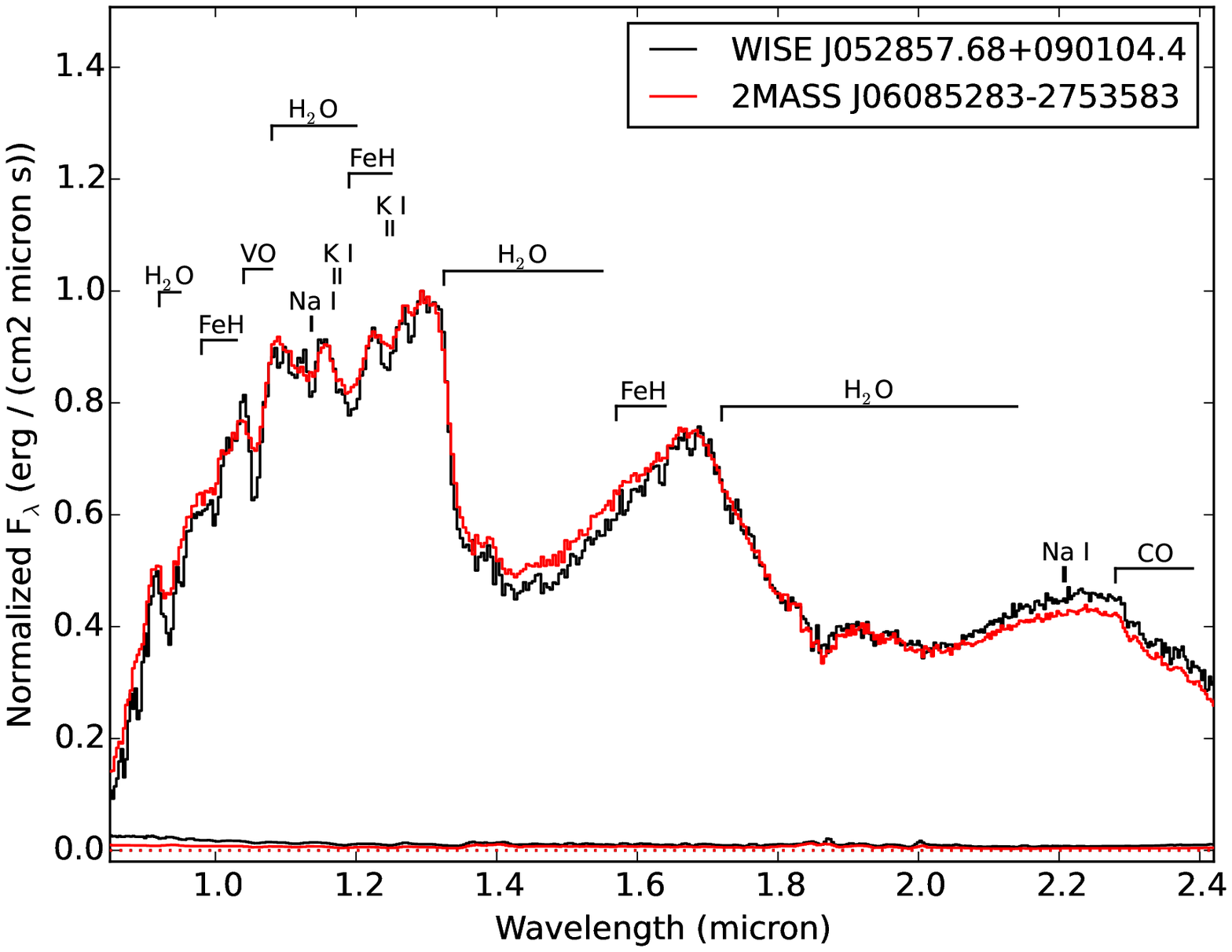}
\caption{IRTF/SpeX near-infrared spectrum of {\namesh} (black line) compared to data for the L1 spectral standard 2MASSW~J2130446$-$084520 (top panel, red line; data from \citealt{2014ApJ...794..143B}) and the young brown  dwarf 2MASS~J0608$-$2753 (bottom panel, red line; data from \citealt{2013ApJ...772...79A}). The spectrum of {\namesh} is normalized at 1.27~$\micron$; the comparison spectra are normalized to their optimal scaling factors (see Eqn~\ref{eqn:alpha}). {Uncertainty spectra are indicated at bottom.} Key absorption features are labeled. \label{fig:comp_spex}}
\end{figure}

We re-examined the classification of {\namesh} by \citet{2013PASP..125..809T} by comparing its SpeX spectrum to the M and L dwarf spectral standards defined in \citet{2010ApJS..190..100K}. Following the prescription of that study, we compared the target spectrum ($T[{\lambda}]$) to standard spectra ($S_k[\lambda]$ for spectral type $k$) over the 0.9--1.4~$\micron$ wavelength range using a $\chi^2$ statistic,
\begin{equation}
\chi^2_k = \sum_{\lambda_i = 0.9~\micron}^{1.4~\micron}\frac{(T[\lambda_i]-\alpha{S_k[\lambda_i]})^2}{\sigma_T[\lambda_i]^2},
\end{equation}
where the denominator includes only the uncertainty of the {\namesh} spectrum. The optimized scale factor is
\begin{equation}
\alpha_k = \left[\sum_{\lambda_i = 0.9~\micron}^{1.4~\micron}\frac{S_k[\lambda_i]T[\lambda_i]}{\sigma_T[\lambda_i]^2}\right]/\left[\sum_{\lambda_i = 0.9~\micron}^{1.4~\micron}\frac{S_k[\lambda_i]^2}{\sigma_T[\lambda_i]^2}\right]
\label{eqn:alpha}
\end{equation}
(cf.\ \citealt{2008ApJ...678.1372C}). The best match was to the L1 standard 2MASSW~J2130446$-$084520 (\citealt{2008ApJ...689.1295K}; Figure~\ref{fig:comp_spex}). However, as originally noted by \citet{2013PASP..125..809T}, there are several distinct peculiarities in the spectrum of {\namesh}, including an unusually peaked $H$-band (1.7~$\mu$m) continuum; weak FeH (1.0~$\mu$m), CO (2.3 $\mu$m) and Na~I (2.2~$\micron$) absorption features; and a prominent 1.05 $\mu$m VO band. All of these features are indicators of low surface gravity \citep{2001MNRAS.326..695L,2003ApJ...593.1074G,2007ApJ...657..511A,2013ApJ...772...79A}.  We computed additional  classifications using the spectral index-based methods of \citet{2001AJ....121.1710R} and \citet{2013ApJ...772...79A}, which yielded types of L1.8$\pm$0.8 and L0.5$\pm$0.5, respectively.\footnote{Uncertainties include propogation of spectrum measurement uncertainties and systematic uncertainty in relations.} The near-infrared spectrum of {\namesh} appears to be that of a peculiar and possibly young L1 dwarf.

We then compared the SpeX spectrum to all 911 optically-classified M5--L5 dwarfs in the SPL. In this case, spectra were compared using the same $\chi^2$ statistic but over the ranges 0.80--1.35~$\micron$, 1.42--1.80~$\micron$ and 1.92--2.45~$\micron$ to avoid regions of strong telluric absorption.   The best-matching spectrum (Figure~\ref{fig:comp_spex}) is that of 2MASS~J06085283$-$2753583 (hereafter 2MASS~J0608$-$2753), a brown dwarf member of the 23$\pm$3~Myr $\beta$ Pictoris Moving Group\footnote{\citet{2014ApJ...783..121G} 
argued that 2MASS~J0608$-$2753 is a candidate member of the 42$^{+6}_{-4}$~Myr Columba association \citep{2011ApJ...732...61Z,2015MNRAS.454..593B}, although the $XYZ$ Galactic coordinates and $UVW$ space velocities are somewhat discrepant. Re-examination of these quantities using the original proper motion in \citet{2010ApJ...715L.165R} make $\beta$ Pictoris a better kinematic match than Columba.}
 \citep{2010ApJ...715L.165R,2014MNRAS.445.2169M}. The spectrum of 2MASS~J0608$-$2753 exhibits the same peculiarities as that of {\namesh}, but with slightly weaker VO and H$_2$O features and a bluer overall spectrum.  

\citet{2008ApJ...689.1295K} reported an optical classification of M8.5$\gamma$ for 2MASS~J0608$-$2753, the suffix indicating very low surface gravity features in the red optical region (see also \citealt{2009AJ....137.3345C}).  We determined equivalent near-infrared gravity classifications for {\namesh} and 2MASS~J0608$-$2753 using the index-based scheme of \citet{2013ApJ...772...79A}, which compares the gravity-sensitive features of FeH, VO, K~I and $H$-band continuum shape.  Both spectra had gravity scores of VLG (very low gravity), consistent with prior analysis of 2MASS~J0608$-$2753 by \citet{2013ApJ...772...79A,2015arXiv150607712G}.  Other VLG-classified brown dwarfs reported in that and subsequent studies (e.g., \citealt{2015ApJ...806..254A,2015arXiv150607712G}) are members of young associations with ages of 5--30 Myr.  \citet{2013ApJ...772...79A} and \citet{2015arXiv150607712G} also classified 2MASS~J0608$-$2753 as an L0 dwarf in the near-infrared, consistent with our classifications based on comparison to standards (L0.5$\pm$0.5) and indices (L0.9$\pm$0.8 for \citealt{2001AJ....121.1710R}).  
It has been previously noted that young brown dwarfs have optical classifications that are up to 3 types earlier than their near-infrared types \citep{2008ApJ...689.1295K}.

\section{Atmospheric Model Fitting}\label{sec:model_fit}

To more quantitatively characterize the physical properties of {\namesh}, we compared the Thompson et al.\ spectrum to  
solar-metallicity BT-Settl atmosphere models \citep{2012RSPTA.370.2765A} over an effective temperature ({\teff}) range of 400--2900~K and a log surface gravity ({\logg}) range of 3.5--5.5 (units of cm~s$^{-2}$). Models were smoothed to the equivalent resolution of the data using a {Hamming filter \citep{BLTJ:BLTJ3874}}, and interpolated from the initial grid (steps of 100~K in {\teff} and 0.5~dex in {\logg}) by linearly interpolating the logarithm of fluxes. We compared the observed spectrum to the models, avoiding the telluric bands, using the same $\chi^2$ statistic as above, and explored parameter space using a custom Markov Chain Monte Carlo (MCMC) code with a Metropolis-Hastings algorithm \citep{1953JChPh..21.1087M,HASTINGS01041970}. We started with an initial {\teff} = 2100 K, based on its L1 near-infrared classification and the {\teff}/spectral-type relations of \citet{2009ApJ...702..154S}, \citet{2013AJ....146..161M}, and \citet{2015ApJ...810..158F}; and an initial {\logg} = 4.5 based on its low surface gravity classification. We then computed a chain of 10$^5$ steps, alternately updating $\vec{\theta}$ = ({\teff}, {\logg}) by random draws from a normal distribution, at each $i^{th}$ parameter step:
\begin{equation}
P(\theta_{(i+1)}|\theta_{(i)}) \propto e^{-\frac{(\theta_{(i+1)} - \theta_{(i)})^2}{2\sigma_{\theta}^2}}
\end{equation}
where $\sigma_{T_{eff}}$ = 50~K and $\sigma_{log\,g}$ = 0.25 dex.  
The criterion to adopt successive parameters was {based on the F-test cumulative distribution function (FCDF),
\begin{equation}
U(0,1) < 1 - FCDF\left(\frac{\chi^2_{(i+1)}}{\chi^2_{(i)}},DOF,DOF\right)
\end{equation} where $U(0,1)$ is a random number drawn from a uniform distribution between 0 and 1 and $DOF$ = 169 is the number of degrees of freedom in the fit, accounting for the resolution of the spectral data and two model parameters\footnote{We had previously employed a pure exponential, $U(0,1) < e^{-0.5(\chi^2_{(i+1)}-\chi^2_{(i)})}$, as a step criterion (e.g., \citealt{2015arXiv151208659A}), but found this to be overly discriminatory due to the large values of $\chi^2$ involved. The FCDF properly accounts for the degrees of freedom in the fit.}}.  We eliminated the first 10\% of the chain and then evaluated the distributions of {\teff} and {\logg} parameters for the remainder.   

\begin{figure}[h]
\epsscale{1.0}
\plotone{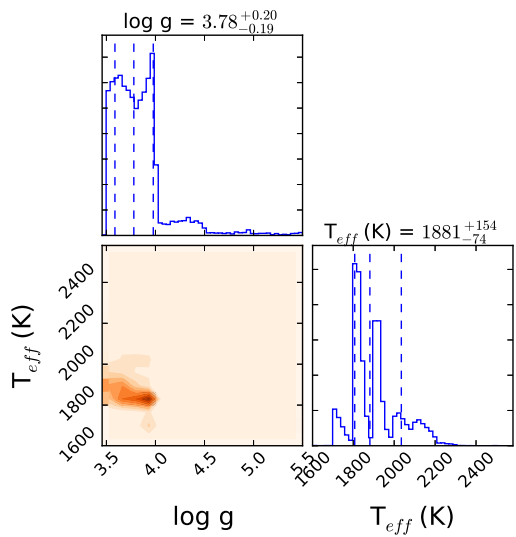}
\caption{{\teff} and {\logg} parameter distributions from our MCMC analysis comparing BT-Settl models to the SpeX data for {\namesh}. {The lower left corner shows the probability density distribution of our two parameters, determined as a percentage of all MCMC chain steps, with color contours ranging from 0\% (light) to 90\% (dark) in steps of 20\%. This distribution indicates a slight negative correlation between the parameters, with higher {\teff}s matching to lower surface gravities.}  
{Marginalized one-dimensional parameter distributions are shown on the wings}, with median and 16\% and 84\% quantiles labeled and listed. Note that structure in the {\teff} distribution arises from our model interpolation scheme. This plot was generated using code by \citet{dan_foreman_mackey_2014_11020}.
\label{fig:mcmc_parameters}}
\end{figure}

\begin{figure}[h]
\epsscale{0.9}
\plotone{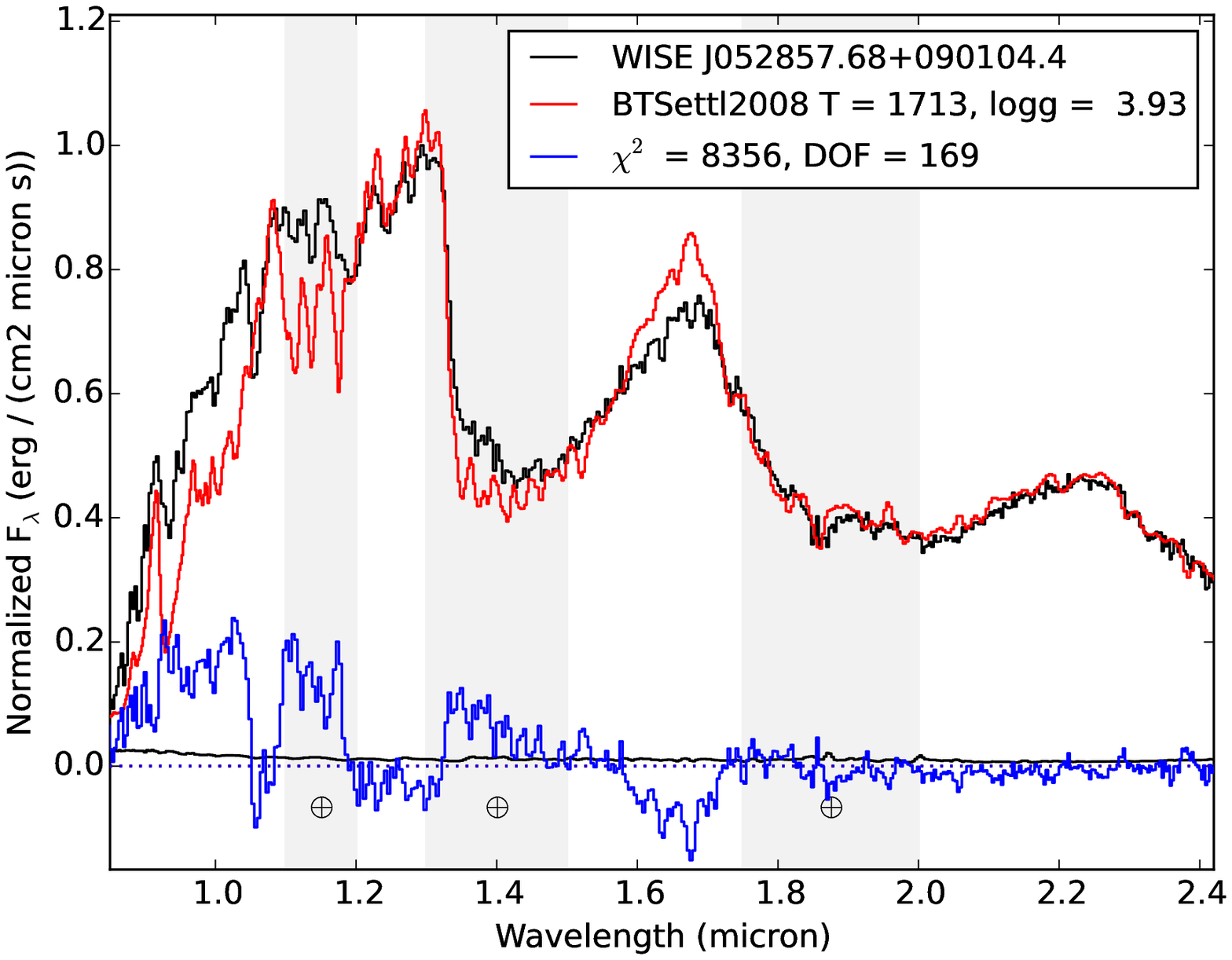} \\
\plotone{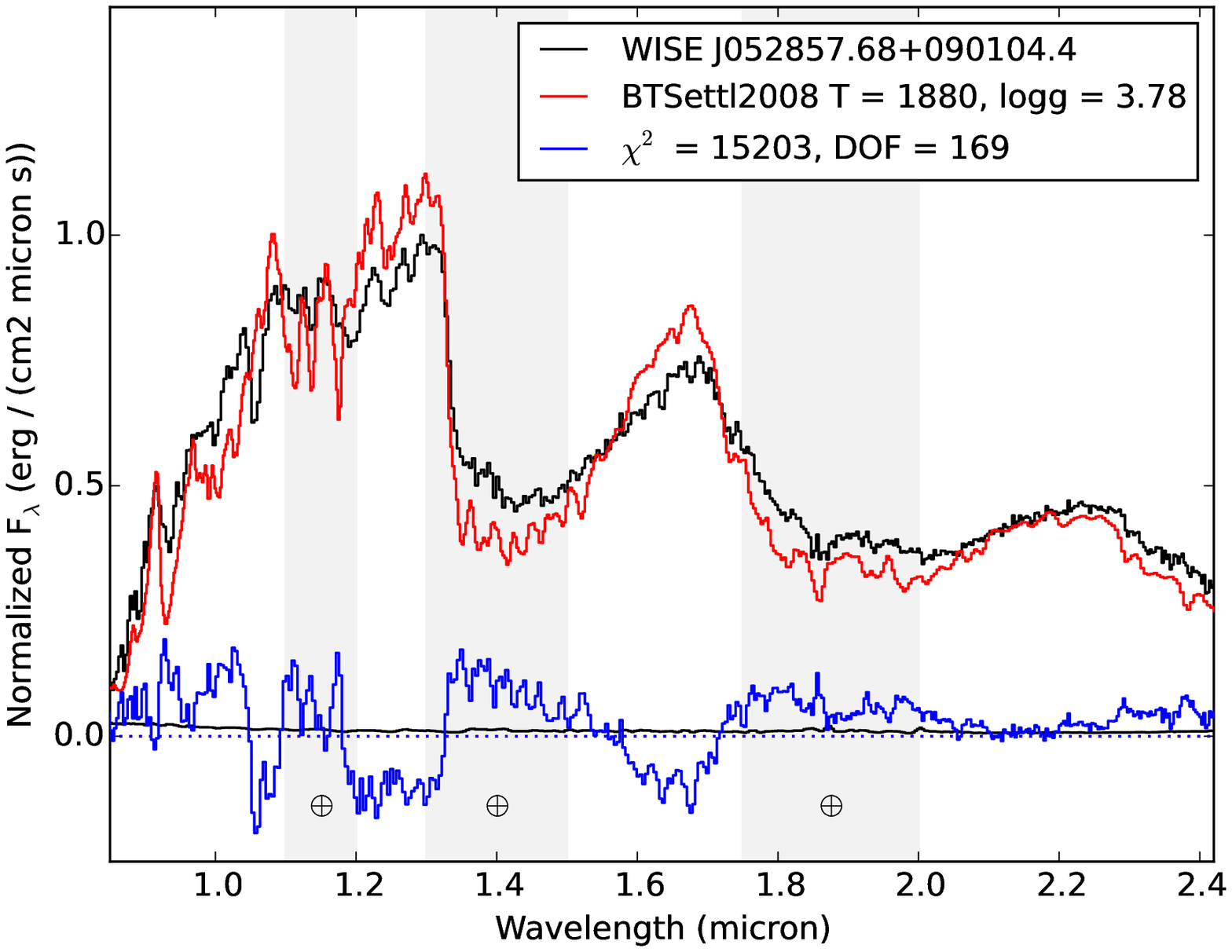}
\caption{SpeX near-infrared spectrum of {\namesh} (black line) compared to the best-fit BT-Settl model from MCMC analysis (top panel) and the model corresponding to the median parameters (bottom panel).  The spectrum of {\namesh} is normalized at 1.27~$\micron$, while the spectra of the atmosphere models (red lines) are normalized to their optimized scaling factors (Eqn.~\ref{eqn:alpha}). The difference spectra ({\namesh} - Model) are shown as blue lines.  The resulting $\chi^2$ values and degrees of freedom (DOF) are listed in the inset boxes.  Regions of strong telluric absorption not included in the fit are indicated by vertical gray bands.
\label{fig:mcmc_bestfit}}
\end{figure}

Table~\ref{tab:mcmc_results} lists the best-fit and median parameter values, while Figure~\ref{fig:mcmc_parameters} displays the distributions of and correlation between {\teff} and {\logg}.
The surface gravity distribution abuts the lower limit of our models, but is constrained to {\logg} $\lesssim$ 4.0, while 
{\teff} is constrained to 1800--2000~K; i.e., about 200~K cooler than the {\teff} estimate based on its spectral type.  
Both distributions exhibit structure due to the model interpolation scheme.
We find median parameters of {\teff} = {\teffresult}~K and {\logg} = {\loggresult}, where the uncertainties correspond to the 16\% and 84\% quantiles of the marginalized parameter distributions.
{The best-fit model is shown in Figure~\ref{fig:mcmc_bestfit}, and has {\teff} = 1713~K and {\logg} = 3.93. 
This model reasonably represents the overall spectral energy distribution of {\namesh}, but shows large deviations around the 1.05~$\micron$ VO band, overly strong H$_2$O at 0.9~$\micron$ and 1.4~$\micron$, and too sharp of an $H$-band peak.
With $\chi^2$ = 8356 for 169 degrees of freedom, this model is clearly not a precise representation of the data.  
The best-fit temperature is also a significant outlier as compared to the inferred {\teff} distribution.
The disagreements between spectral models and data necessitate some skepticism in the parameters inferred from the best-fit model.}

\begin{deluxetable}{lcc}
\tabletypesize{\small}
\tablecaption{Results from MCMC Model Fitting Analysis\label{tab:mcmc_results}}
\tablewidth{0pc}
\tablehead {
\colhead{Parameter}	& \colhead{Best-Fit} & \colhead{Median Value\tablenotemark{a}} \\
}
\startdata
{\teff} (K) & 1713 &  {\teffresult} \\
{\logg} (K) & 3.93  &  {\loggresult} \\
Age (Myr)\tablenotemark{a} & 21 & 18$^{+4}_{-17}$   \\
Mass ({\mjup})\tablenotemark{a} & 13 & 13$^{+3}_{-6}$ \\
Radius ({\rsun})\tablenotemark{a} & 0.15 & 0.18$^{+0.03}_{-0.02}$ \\
$\chi^2$ (DOF) & 8356 (169)  & \nodata \\
\enddata
\tablenotetext{a}{Based on the {\teff} and {\logg} parameter distributions and evolutionary models of \citet{2003A&A...402..701B}.}
\end{deluxetable}

{We attempted to infer physical parameters (mass and age) from our model-fit parameters and the evolutionary models of \citet{2003A&A...402..701B}.  This proved to be challenging as the model parameters span an epoch (10--50~Myr) of active deuterium fusion in low-mass brown dwarfs (15--30~{\mjup}), {sparsingly sampled in the evolutionary models}, during which the thermal pressure  briefly reverses the general trend of decreasing surface gravity with decreasing mass {via radius expansion} \citep{2001RvMP...73..719B,2003A&A...402..701B,2011ApJ...727...57S}.
Interpolation over this feature produces large uncertainties in the inferred age ($\tau$ =  18$^{+4}_{-17}$~Myr) and mass\footnote{1 {\mjup} = 1 jovian mass =  0.000955~{\msun}. {Reported uncertainties on mass and age again reflect the 16\% and 84\% quantiles.}} (M = 13$^{+3}_{-6}$~{\mjup}) of {\namesh}. Nevertheless, the age is similar to that of $\beta$ Pictoris and, as discussed below, 32~Orionis, while the mass straddles the deuterium-burning boundary.  We re-examine the physical properties of {\namesh} in Section~\ref{sec:discussion}.}

\section{Association of {\namesh} with 32 Orionis}\label{sec:association}

{\namesh} is spatially aligned with several young associations and star forming regions in the general direction of Orion, although it is too nearby to be a member of the any of the Orion OB1 subgroups; e.g., $\lambda$ Orionis or Ori OB1a \citep{1989ARA&A..27...41G}. Comparing its apparent 2MASS magnitudes to the absolute magnitude scale of \citet{2012ApJS..201...19D} for spectral types of M9 and L1 (which encompass both its near-infrared and likely optical classifications; see Section~\ref{sec:spex_analysis}), we find mean distances of 96$\pm$11~pc to 74$\pm$8~pc, respectively, or a combined average of 81$\pm$13~pc (Table~\ref{tab:distance}).  
We also computed the corresponding distances for the WISE $W2$ band and found these to be 30\% smaller, suggesting excess flux at 5~$\micron$. Indeed, the $J-W2$ = 2.62$\pm$0.12 color for {\namesh} is unusually red for M9--L1 dwarfs, even among young sources \citep{2012ApJ...752...56F,2015ApJ...810..158F}.  This feature is discussed further in the following section.
{Note that these distances are likely underestimated, as the \citet{2012ApJS..201...19D} relation is defined for evolved field dwarfs. Young late-M and L dwarfs (10--100~Myr) are generally found to be overluminous in these bands, due to their larger radii and spectral classification offsets \citep{2012ApJS..201...19D,2012ApJ...752...56F,2015ApJ...810..158F}.}

\begin{figure*}[t]
\epsscale{1.1}
\plottwo{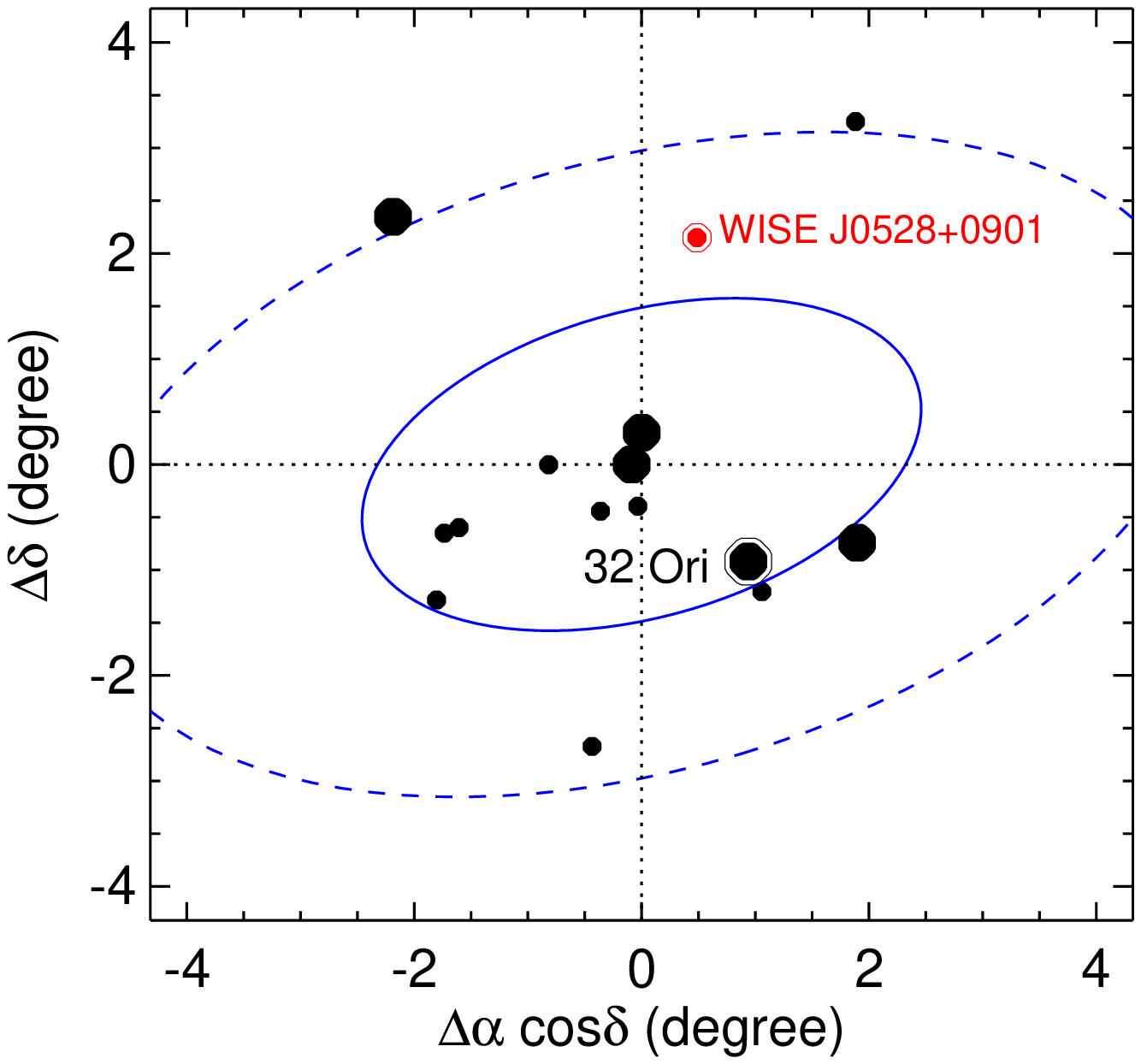}{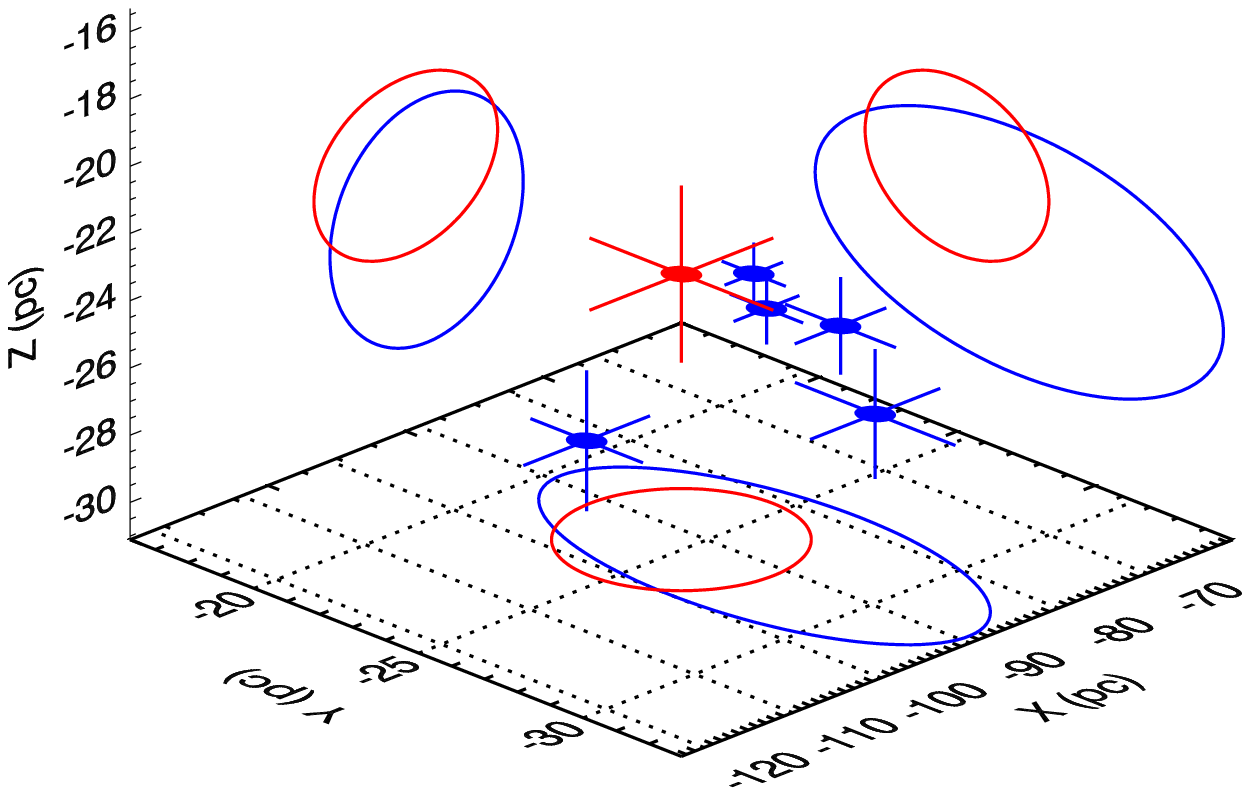}
\caption{(Left) Sky positions of 32 Orionis members listed in \citet[black points]{2015MNRAS.454..593B} and {\namesh} (red point), relative to the mean sky position of the former.  Large symbols indicate the five 32 Ori members with parallax measurements. 1$\sigma$ (solid line) and 2$\sigma$ (dashed line) dispersion ellipses for group members are indicated. (Right) Galactic $XYZ$ coordinates of the 32 Orionis members with parallax measurements (blue points) compared to {\namesh}, assuming a distance of 98$\pm$11~pc; i.e., consistent with an M9 spectral type. The projected 3$\sigma$ dispersion (blue lines) and error ellipse for {\namesh} (red line) are indicated. Note that the coordinate system shown is centered on the Sun.
\label{fig:32ori}}
\end{figure*}

\begin{deluxetable*}{lccccc}
\tabletypesize{\small}
\tablecaption{Distance Estimates for {\namesh}\tablenotemark{a}\label{tab:distance}}
\tablewidth{0pc}
\tablehead {
& & \multicolumn{2}{c}{M9} &  \multicolumn{2}{c}{L1} \\
\cline{3-4} \cline{5-6}
\colhead{Filter}	& \colhead{Apparent Mag} & \colhead{Absolute Mag} & \colhead{d (pc)} & \colhead{Absolute Mag} & \colhead{d (pc)} }
\startdata
2MASS $J$ 	& 16.26$\pm$0.12	& 11.3$\pm$0.4	& 102$\pm$21	& 12.0$\pm$0.4	& 72$\pm$12	\\
2MASS $H$	& 15.44$\pm$0.13	& 10.6$\pm$0.4	& 93$\pm$18	& 11.1$\pm$0.4	& 77$\pm$15	\\
2MASS $K_s$	& 14.97$\pm$0.12	& 10.1$\pm$0.4	& 94$\pm$18	& 10.7$\pm$0.4	& 75$\pm$15	\\
WISE $W2$	& 13.64$\pm$0.04	& 9.6$\pm$0.4	& 65$\pm$11	& 10.1$\pm$0.4	& 52$\pm$7	\\
\hline
SpT Average\tablenotemark{b} & 	&				& 96$\pm$11	&		& 74$\pm$8 \\
Overall Average\tablenotemark{b}  & 	&			& 81$\pm$13 \\
{32 Orionis\tablenotemark{c}}  & 	&			& 93$\pm$5 \\
\enddata
\tablenotetext{a}{Based on the absolute magnitude/spectral type relations of \citet{2012ApJS..201...19D}.}
\tablenotetext{b}{Combining only 2MASS $JHK_s$ distances.}
\tablenotetext{c}{Based on the HIPPARCOS parallaxes of five 32 Orionis members \citep{2007A&A...474..653V}.}
\end{deluxetable*}

Within the uncertainties, {\namesh} is roughly at the distance of the B5+B7 star 32 Orionis (93$^{+6}_{-5}$~pc, \citealt{2007A&A...474..653V}), separated by only 3$\fdg$1 on the sky. \citet{2007IAUS..237..442M} first identified a co-moving group of X-ray-bright T Tauri stars around 32 Orionis; and  \citet{2015MNRAS.454..593B} have analyzed the kinematics, disc fraction and age diagnostics of 20 co-moving members, determining an isochronal age of 24$^{+4}_{-3}$~Myr.  \citet{2013A&A...558A..53K} have also identified this system, characterizing it as an open cluster with an isochronal age of 32~Myr.
{The system exhibits modest reddening, E(B-V) = 0.04$\pm$0.02.}
Figure~\ref{fig:32ori} compares the equatorial sky coordinates of 32 Orionis members\footnote{\citet{2015A&A...584A..26B}
have recently proposed the bright star Bellatrix ($\gamma$ Orionis) as a member of the 32 Orionis group based on its position and age; however, the proper motion of this source is sufficiently discrepant from the mean of the other members that its 3D velocity differs by 12~{\kms}, a 10$\sigma$ discrepany. We have chosen not to include this source in our comparative analysis until the membership of the naked eye star can be confirmed.} to those of {\namesh}.
While slightly outside the ``core'' of the group, {\namesh} is still within the 2$\sigma$ dispersion ellipse of members on the sky. We also compared the Galactic $XYZ$ positions of the five 32 Orionis members with {HIPPARCOS} parallaxes (32~Orionis, HD~34500, HD~35656, HD~35714 and HD~36823; \citealt{2007A&A...474..653V}) to that of {\namesh}, assuming its M9 dwarf distance. {\namesh} is spatially coincident with the member stars, overlapping a 3$\sigma$ dispersion sphere in all three dimensions.  There is therefore strong spatial evidence that {\namesh} is a member of 32 Orionis.

With regard to kinematics, \citet{2007IAUS..237..442M} report a mean $\vec{\mu}$ = (+7,$-$33)~mas~yr$^{-1}$ for group members. We performed a new cross-match of 2MASS\footnote{\citet{2013PASP..125..809T} did not make use of 2MASS astrometry for {\namesh} as it appears only in the 2MASS Reject Table \citep{2003yCat.2246....0C} due to a mis-labled artifact flag in the $J$-band.} and WISE sources in the vicinity of {\namesh} and found $\vec{\mu}$ = ($-$11$\pm$10, $-$39$\pm$12)~mas~yr$^{-1}$ for this source, inconsistent with the zero motion reported by  \citet{2013PASP..125..809T} but consistent with the mean motion of group members. 
32 Orionis itself has a heliocentric RV of +18.6$\pm$1.2~{\kms} \citep{2007AN....328..889K}, which is identical to the RV measured here for {\namesh}.  

\section{Discussion}\label{sec:discussion}

The observed and estimated physical properties of {\namesh} are summarized in Table~\ref{tab:properties}.  Spectral, spatial and kinematic evidence are mutually consistent with {\namesh} being a young substellar object and member of the 32 Orionis group. The isochronal age of this system, 24$^{+4}_{-3}$~Myr \citep{2015MNRAS.454..593B}, is similar to the age inferred from spectral and evolutionary model analysis, as well as that of {\namesh}'s $\beta$ Pictoris counterpart, 2MASS~J0608$-$2753. At this age, the estimated mass from the \citet{2003A&A...402..701B} evolutionary models, assuming a conservative estimate of {\teff} = 1900$\pm$250~K {(which encompasses the temperature range inferred from our model fits)}, is 14$^{+4}_{-3}$~{\mjup}, {again} overlapping the deuterium-burning limit.  The estimated {\logg} from the evolutionary models is 4.16$^{+0.04}_{-0.08}$, only 1.3$\sigma$ higher than the atmospheric model fits. 
{We also examined predictions from the \citet{2008ApJ...689.1327S} evolutionary models which include cloud opacity effects in brown dwarf thermal evolution, and found identical values for mass (M = 14$^{+2}_{-2}$~{\mjup}) and surface gravity ({\logg} = 4.12$^{+0.05}_{-0.10}$), whether or not cloud opacity is included.}
There is therefore good agreement between spectral and evolutionary model analyses for this source, assuming it is a 32 Orionis group member.

\begin{deluxetable}{lcc}
\tabletypesize{\small}
\tablecaption{Summary of Properties of {\namesh}\label{tab:properties}}
\tablewidth{0pc}
\tablehead {
\colhead{Parameter}	& \colhead{Value} & \colhead{Reference} \\
}
\startdata
$\alpha$ (J2000)\tablenotemark{a} &	$05^h28^m57\fs68$ & 1 \\
$\delta$ (J2000)\tablenotemark{a}   & +$09^{\circ}01'04\farcs4$			&1	\\
NIR Spectral Type						&L1 VLG							&2	\\
2MASS $J$									&16.26$\pm$0.11								&1	\\
2MASS $H$									&15.44$\pm$0.12								&1	\\
2MASS $K_s$									&14.97$\pm$0.11								&1	\\
WISE $W1$									&14.21$\pm$0.03								&3	\\
WISE $W2$									&13.64$\pm$0.04								&3	\\
$\mu_{\alpha}\cos(\delta)$ (mas yr$^{-1}$)	&$-$11$\pm$10 & 2 \\
$\mu_{\delta}$ (mas yr$^{-1}$) & $-$39$\pm$12	&2	\\
d (pc)\tablenotemark{b}			&93$\pm$5							&2,4	\\
RV ({\kms})								&18$\pm$4						&2	\\
$U$ ({\kms})\tablenotemark{b} & $-$11$\pm$4 & 2 \\
$V$ ({\kms})\tablenotemark{b} & $-$15$\pm$5 & 2 \\
$W$ ({\kms})\tablenotemark{b} & $-$17$\pm$5 & 2 \\
{\teff}	 (K) &{\teffresult}		&2	\\
{\logg} (cm~s$^{-2}$) & {\loggresult}		&2	\\
Age (Myr)\tablenotemark{b}									& 24$^{+4}_{-3}$					&5	\\
Mass ({\mjup})\tablenotemark{b}								& 14$^{+4}_{-3}$			&2,6	\\
\enddata
\tablenotetext{a}{Julian Date Epoch 2451569.7 (2000.07).}
\tablenotetext{b}{Assuming membership in the 32 Orionis group.}
\tablerefs{(1) 2MASS \citep{2006AJ....131.1163S}; (2) This work; (3) WISE \citep{2010AJ....140.1868W}; (4) {HIPPARCOS} \citep{2007A&A...474..653V}; (5) \citet{2015MNRAS.454..593B}; (6) \citet{2003A&A...402..701B}.}
\end{deluxetable}

\begin{figure}[h]
\epsscale{0.9}
\plotone{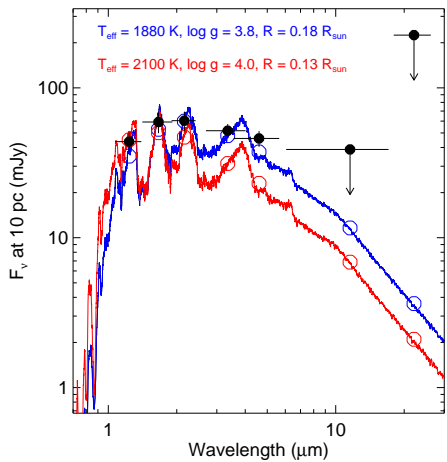}
\caption{Absolute 2MASS and WISE photometric fluxes for {\namesh} (black points) assuming a distance of 93$\pm$5~pc. The two longest wavelength points are upper limits based on WISE non-detections.
{These measured fluxes are compared to two BT-Settl spectral models: 
one with parameters inferred from the L1 VLG classification of the source ({\teff} = 2100~K, {\logg} = 4.0; {red lines and circles}),  and one with the median parameters from spectral model fitting ({\teff} = 1880~K, {\logg} = 3.8; {blue lines and circles}).} 
All models are scaled to agree with absolute 2MASS photometry, and the corresponding radii for those scale factors are listed. 
\label{fig:sed}}
\end{figure}

The additional agreement in the spectral peculiarities between {\namesh} and its similarly-aged counterpart 2MASS~J0608$-$2753 supports the conjecture that these features are consistent metrics of age. {Nevertheless, {\namesh} itself appears to be unusually red in $J-W2$ color as compared to its young counterparts, in excess of cluster reddening.}
Condensate cloud effects may play an important role in these color differences \citep{2013ApJ...772...79A,2013AJ....145....2F,2015ApJ...799..203G}, but so can circumstellar structure: disks and planetary companions \citep{2010AJ....140.1486L,2010ApJ...714...45L,2012ApJ...757..163S}.  On the other hand, \citet{2015ApJ...810..158F} have shown that young L dwarfs can be up to 300~K cooler than equivalently (optically) classified field sources, and that this lower temperature provides an explanation for the red $J-W2$ colors. 
{For {\namesh}, we find evidence that this last explanation is the most likely.  
Our model-fit temperature is about $\approx$200~K cooler than other L1 field dwarfs, and our median-parameter 
spectral model is also a better match to the overall near-infrared spectral energy distribution (SED) of {\namesh} than a warmer model (Figure~\ref{fig:sed}). 
The median-parameter SED ($\chi^2$ = 11.1 for the 2MASS $JHK_s$ and WISE $W1$ and $W2$ bands) requires no significant 3--5~$\micron$ excess, and the scale factor to align this model with the photometry corresponds to a radius of 0.18$\pm$0.02~{\rsun}, which is consistent with predictions for the radius of a {\teff} = 1880~K, 25~Myr-old brown dwarf from evolutionary models (0.16~{\rsun} for \citealt{2003A&A...402..701B}).
In contrast, a very low gravity L1 dwarf model ({\teff} = 2100~K, {\logg} = 4.0) requires significant reddening to match the data, and a radius much smaller (0.13$\pm$0.02~{\rsun}) than predicted by evolutionary models (0.17~{\rsun}).} 
Given the consistency of our spectral and evolutionary modeling analyses {for our median-fit parameters}, we conclude that {{\namesh} is cooler than equivalently classified field L dwarfs, and furthermore} shows no evidence of warm circumstellar material, although these data cannot rule out the presence of a cooler ($\lesssim$300~K) disk {or companion}. No {warmer} companions were {resolved} in laser guide star adaptive optics observations reported in \citet{2015AJ....150..163B} to a limit of $\Delta{H}$ = 3 at 0$\farcs$2 (19~AU).

A more detailed analysis of the broad-band spectral properties of {\namesh}, including optical spectroscopy, is needed to assess whether other aspects of this source, such as magnetic emission, are remarkable.
{\namesh} joins a growing list of young, very low-mass brown dwarfs and planetary-mass objects whose ages, distances, kinematics and compositions (assuming a common initial reservoir of gas and dust) are in common with their more massive siblings. As demonstrated here, these systems provide important validation tests for atmospheric and evolutionary models, while also statistically probing the origins of brown dwarfs in our Galaxy.

\acknowledgments
The authors thank Jorge Araya at Magellan Observatory for his assistance with the observations, and
Richard Faherty for providing a facility to complete this work.
We also thank our referee, Derek Homeier, for his prompt and helpful review.
The material presented here is based in part on work supported by the National Aeronautics and Space Administration under Grant No.\ NNX15AI75G, and funding from the McNair Scholars Program.
EEM acknowledges support from NSF grant AST-1313029.
This research has made use of the SIMBAD database,
operated at CDS, Strasbourg, France;
NASA's Astrophysics Data System Bibliographic Services;
the M, L, T, and Y dwarf compendium housed at DwarfArchives.org;
and the SpeX Prism Libraries housed at \url{http://www.browndwarfs.org/spexprism}.
The authors recognize and acknowledge the very significant cultural role and reverence that the summit of Mauna Kea has always had within the indigenous Hawaiian community.  We are most fortunate and grateful to have the opportunity to conduct observations from this mountain.

{\it Facilities:} \facility{Magellan (FIRE)},  \facility{IRTF (SpeX)}

\end{document}